\documentstyle[pra,preprint,epsfig,aps]{revtex}
\begin{document}
\draft
\title{Dressed Bose-Einstein Condensates in High-Q Cavities}
\author{Elena V. Goldstein, Ewan M. Wright and P. Meystre}
\address{Optical Sciences Center, University of Arizona, Tucson, AZ 85721}
\maketitle
\begin{abstract}
We propose and analyze a way in which effective multicomponent condensates
can be created inside high-Q multimode cavities. In contrast to the situation
involving several atomic species or levels, the coupling between the various
components of the dressed condensates is linear. We predict analytically and 
numerically confirm the onset of instabilities in the quasiparticle 
excitation spectrum.
\end{abstract}
\date{today}
\section{Introduction}
The Bose-Einstein condensation of low density atomic samples 
\cite{AndEnsMat95,DavMewAnd95} provides one 
with a new paradigm in many-body theory, atomic physics, quantum optics, 
and nonlinear dynamics. Below the critical temperature, condensates are
described to an excellent degree of accuracy by a scalar 
nonlinear Schr\"odinger equation, the Gross-Pitaevskii equation 
describing the dynamics of the condensate wave function 
\cite{LifPit80,EdwBur95,RupHolBurEdw95,BayPet96}. The elementary
excitations of the condensate evaluated from a Bogoliubov linearization about
the condensate solution are in good qualitative agreement with experiments
\cite{Fet96,Str96,SinRok96,EdwRupBurDodCla96,JinEnsMat96,MewAndDruKur96,Gri96,KagShlWal96}, and
so is the Hartree mean-field energy of the system \cite{EnsJinMat96,MewAndDru96}.

Nonlinear Schr\"odinger equations are of course ubiquitous in physics, and
have been studied in great detail in the past, in situations from fluid 
dynamics \cite{NewPasLeg93}
to phenomenological models of field theories \cite{Col85} and to nonlinear optics
\cite{NewMol92}.
They play an important role in the study of pattern formation in beam
propagation \cite{LugNas94}, and their soliton solutions 
find applications in problems
such as light propagation in fibers \cite{Aga95}. 
From this work, it is known that the
dynamical and stability properties of multicomponent nonlinear Schr\"odinger 
equations can be vastly different from those of their scalar versions, and 
lead to a wealth of new effects \cite{Aga95,EilLomSco85,KinLut90,GolMey97}. 
It would be of considerable interest
to generalize these ideas to the case of matter waves and to have  
multicomponent condensates available.

While it is not generally possible to create two coexisting condensates
inside a trap, exceptions are possible, as recently demonstrated in 
Rubidium experiments by the JILA group \cite{MyaBurGhr97}. 
However, this coexistence relies on
a fortuitous coincidence of the scattering lengths for the two Zeeman
sublevels involved \cite{JulMieTieWil97}, 
a coincidence that cannot be generally counted on. There
are proposals to optically change the s-wave scattering length of ground 
state atoms \cite{FedKagShlWal96}, but whether this can be used 
to produce coexisting condensates remains to be seen.  The goal of the present 
paper is to propose and analyze a method by
which effective multicomponent condensates can be generated inside a high-Q
multimode optical resonator. The cavity photons dress the condensate, very much
like atoms can be dressed by electromagnetic fields \cite{CohDupGry92}, 
and the various dressed
condensate states are coupled, e.g. via an electric dipole interaction. 
Hence, the condensate inside the cavity should be thought of as a 
coupled multicomponent system, each component subject to a nonlinear equation,
and in addition coupled to its neighboring components. In contrast to the 
situation involving two (or more) atomic species or levels, the coupling 
between the
various components of the dressed condensate is linear, rather than resulting
from collisions and hence nonlinear. Nonetheless, we submit that this method
permits to generate and study "coupled condensates" in a controlable --- and 
at least in principle simple --- way.

This paper is organized as follows: Section II defines our model and uses
a Hartree variational principle to derive coupled nonlinear Schr\"odinger 
equations describing the evolution of a dressed condensate in a two-mode 
cavity. Section III specializes to the case where only one photon is present
inside the resonator, and thus only two dressed condensate components
are of importance. In that case, the problem reduces to the so-called
discrete self-trapping equations for a dimer familiar in nonlinear physics.
These equations are integrable in free space, but not in the trap situation
that we consider here. We solve them approximately in the Thomas-Fermi 
approximation and study the spectrum of elementary excitations. We predict the
onset of instabilities in the system, and compare these analytical results 
with an exact numerical solution of the equations. Finally, Section IV is 
a summary and conclusion.

\section{Physical Model}
\subsection{Basic theory}
Our model system comprises a Bose-Einstein condensate
(BEC) which interacts with two counter-propagating modes
supported by a high-Q ring cavity.  We assume a cavity QED configuration 
\cite{Har92}
and neglect all field modes except the two of interest, so that the
electric field operator can be written as
\begin{equation}
{\bf E}({\bf r},t) = i
\hat{\mbox{\boldmath {$\epsilon$}}}
{\cal E}_p\left [ a_1 e^{ikz} + a_2 e^{-ikz} \right ]
e^{-i\omega_c t} + h.c.  ,
\end{equation}
where $\hat{\mbox{\boldmath {$\epsilon$}}}$ 
is the unit polarization vector of the light,
${\cal E}_p=\sqrt{\hbar\omega_c/2\epsilon_0 V}$ is the electric field
per photon for light of frequency $\omega_c$ in a mode volume $V$,
$k=\omega_c/c=2\pi/\lambda_c$ the light wavevector,
and $a_{1,2}$, and $a_{1,2}^\dagger$
are annihilation and creation operators of the cavity modes satisfying Bose 
commutation relations $[a_i,a_j^\dagger]=\delta_{ij}$.
For this treatment we neglect the detailed mode structure of 
the field in the transverse plane perpendicular to the optical-axis 
$z$, assuming that it is homogeneous on the spatial scale of the BEC, 
and that it is unaffected by its presence.

The atoms comprising the condensate are confined by an external trapping
potential $U({\bf r})$ which binds the atoms on a sub-wavelength
scale along the longitudinal axis $z$.
In addition they interact with the cavity field, which induces
transitions between the ground and excited electronic states.
The single-particle Hamiltonian $H_0$ for the atoms,
in an interaction picture with the optical frequency removed, then reads
\cite{MeySar91}
\begin{equation}
H_0=\frac{{\bf p}^2}{2m}+U({\bf r})+
\hbar\delta\sigma_+ \sigma_-+\hbar \Omega_0[\sigma_+(a_1+a_2)+
\sigma_-(a_1^\dagger+a_2^\dagger)]  ,
\label{hamsin}
\end{equation}
where we have located the BEC at $z=0$ without loss of generality,
${\bf p}$ is the center-of-mass atomic momentum, $m$ the atomic mass,
$\delta\equiv\omega_a-\omega_c$ is atom-field detuning, $\omega_a$ 
being the atomic transition frequency, $\Omega_0=d{\cal E}_p/\hbar$ is 
the strength of
the atom-field coupling, $d$ being the atomic dipole-matrix element, and
$\sigma_+$, $\sigma_-$ are pseudo-spin
atomic raising and lowering operators for transitions between
the ground and excited atomic states.
We consider the case of large atom-field detuning for which the
excited atomic state can be adiabatically eliminated. This results in  the
following effective single-particle Hamiltonian (see Appendix A for details)
involving only the ground atomic state
\begin{equation}
H_{eff}=\frac{{\bf p}^2}{2m}+U({\bf r})+
\frac{\hbar \Omega_0^2}{\delta}\left(a_1^\dagger a_1+
a_2^\dagger a_2+a_2^\dagger a_1+a_1^\dagger a_2\right)  .
\label{hamsineff}
\end{equation}
The four terms involving field mode operators in this effective Hamiltonian
describe virtual transitions involving the absorption of a photon 
from mode 1 followed by
re-emission into mode 1, the same but for mode 2, absorption of a photon from
mode 1 followed by re-emission into mode 2, and vice versa.
The last two of these processes are allowed since the length $\delta z$ of the
BEC is taken to be less than an optical wavelength, yielding an associated 
momentum uncertainty $\delta p_z \approx\hbar/\delta z > 2\hbar k$. Hence the
momentum deficit involved in the transfer of a photon from one direction 
to the other around the cavity is within the Heisenberg uncertainty principle.
We also note that the effective Hamiltonian conserves the total number of 
photons $n=(a_1^\dagger a_1+a_2^\dagger a_2)$ which is then
a good quantum number.

Proceeding now to include many-body interactions, the second-quantized 
Hamiltonian describing our system is \cite{LifPit80}
\begin{equation}
{\cal H} =
\int d1 d2 \langle 1|H_{eff}|2\rangle \Psi^\dagger(1) \Psi(2)
+ \frac{1}{2}\int d\{\ell\} \langle 1,2|V|3,4\rangle
\Psi^\dagger(1) \Psi^\dagger(2) \Psi(3) \Psi(4)  ,
\label{hnh}
\end{equation}
where $\ell$ denotes a full set of quantum numbers, and
$\Psi(\ell)$ and $\Psi^\dagger(\ell)$ are the usual atomic field
annihilation and creation operators, which for bosonic atoms satisfy
the commutation relations
\begin{equation}
[\Psi(\ell),\Psi^\dagger(\ell ')] = \delta(\ell-\ell ') .
\label{psicom}
\end{equation}
The two-body potential is in the limit of s-wave scattering \cite{LifPit80}
\begin{equation}
V({\bf r}_1,{\bf r}_2)=\hbar V_0 \delta({\bf r}_1-{\bf r}_2),
\end{equation}
where $V_{0}=4\pi\hbar a/m$ measures the strength of the
two-body interaction, $a$ being the s-wave
scattering length. Here we consider a repulsive interaction
so that $a>0, V_0>0$.
\subsection{Coupled-condensate equations}
The second-quantized Hamiltonian for our system conserves both the number
of atoms and the total number of photons, so we consider a state comprising
$N$ atoms and $n$ photons.  The state of the system can be written in the form
\begin{equation}
|\Psi_{N,n}(t) \rangle =\sum_{n_1+n_2=n}\int{d{\bf r}_1}
\ldots\int{d{\bf r}_N}f_{n_1,n_2}
({\bf r}_1,\ldots,{\bf r}_N,t)
\Psi^\dagger({\bf r}_1)\ldots\Psi^\dagger({\bf r}_N)|0,n_1,n_2 \rangle  ,
\end{equation}
where the summation runs over all positive integers $n_{1,2}$ obeying
$n_1+n_2=n$, $f_{n_1,n_2}({\bf r}_1,\ldots,{\bf r}_N,t)$ is the many-particle
Schr\"odinger wave function for the BEC
given there are $n_1$ photons in mode 1
and $n_2$ photons in mode 2, and $|0,n_1,n_2 \rangle$ is the state with
no ground state atoms present, $n_1$ photons in
mode 1 and $n_2$ photons in mode 2. To proceed we invoke the 
Hartree approximation, which is appropriate
for a Bose condensed system in which the atoms are predominantly in the same
state.  The Hartree approximation is therefore strictly valid at zero temperature,
and for a weakly interacting Bose gas as assumed here, so that the condensate fraction
is close to unity \cite{LifPit80}.  Accordingly the many-particle wave function
is written as a product of Hartree wave functions
\begin{equation}
f_{n_1,n_2}({\bf r}_1,\ldots,{\bf r}_N,t)=
\prod_{i=1}^{N}\phi_{n_1,n_2} ({\bf r}_i,t)  .
\end{equation}
Here $\phi_{n_1,n_2} ({\bf r},t)$ is the Hartree wave function which represents
the state the atoms occupy.  The equation of motion for
$\phi_{n_1,n_2}({\bf r},t)$ results from the Hartree variational principle
\cite{Neg82}
\begin{equation}
\frac{\delta}{\delta\phi_{n_1,n_2}^*}
\left [\langle \Psi_{N,n}(t)|i\hbar\frac{\partial}{\partial t} - {\cal H}
|\Psi_{N,n}(t)\rangle \right ] = 0  ,
\end{equation}
and takes the form of a
system of coupled nonlinear Schr\"odinger equations, or Gross-Pitaevskii 
equations
\begin{eqnarray}
i\hbar\dot
\phi_{n_1,n_2} ({\bf r},t)&=&
\left[ -\frac{\hbar^2}{2m}
\mbox{\boldmath {$\nabla$}}^2 + U({\bf r}) \right] \phi_{n_1,n_2}({\bf r},t)
+\hbar V_0 N|\phi_{n_1,n_2}({\bf r},t)|^2 \phi_{n_1,n_2}({\bf r},t)
\nonumber\\
&+& \frac {\hbar{\Omega_0}^2}{\delta}
\left( \sqrt {n_1(n_2+1)}
\phi_{n_1-1,n_2+1}({\bf r},t)+
\sqrt{(n_1+1)n_2} 
\phi_{n_1+1,n_2-1}({\bf r},t)\right)  ,
\label{sys0}
\end{eqnarray}
where the photon numbers $n_{1,2}$ in modes 1 and 2 again run over 
all positive integers obeying $n_1+n_2=n$.
In the limit $\Omega_0=0$ there is no coupling between the cavity 
modes and the BEC and Eq. (\ref{sys0}) is the usual scalar Gross-Pitaevskii 
equation for the condensate.  In contrast, for non-zero values of 
$\Omega_0$ the processes involving absorption of a photon 
from one direction and re-emission into the other direction
lead to a linear coupling between the state with $(n_1,n_2)$ photons and those
with $(n_1-1,n_2+1)$ and $(n_1+1,n_2-1)$ photons, the notation 
$(n_1,n_2)$ meaning $n_1$ photons in state 1 and $n_2$ photons in state 2. As 
a result, the system is generally a superposition of states with different
$(n_1,n_2)$.

As they stand, Eqs. (\ref{sys0}) account for the full three-dimensional
structure of the BEC.  In order to make our presentation as straightforward
as possible we now make some further simplifying assumptions, but we stress that
these are not essential and do not limit the generality of the conclusions
that we draw.  To proceed we write the trapping potential
explicitly as \cite{EdwBur95,RupHolBurEdw95,BayPet96}
\begin{equation}
U({\bf r})=\frac{m\omega _{\perp }^{2}}{2}
\left({\bf r}_{\perp} ^{2}+\lambda^{2}z^{2} \right),
\label{trapot}
\end{equation}
thereby separating the longitudinal potential out from the transverse
trapping potential.  Here
${\bf r}=({\bf r}_{\perp },z),{\bf r}_{\perp }$
being the transverse position coordinate,
$\omega _{\perp }$ the transverse angular
frequency of the trap, and $\lambda =\omega _{z}/\omega _{\perp }$
is the ratio of the longitudinal to transverse frequencies.
Here we assume that $\lambda \ll 1$
so that the longitudinal trapping is
much weaker than the transverse trapping, hence giving the BEC density profile
a cigar structure \cite{MewAndDru96}.\footnote{The opposite case of
weak transverse trapping $\lambda\gg 1$ corresponds to the BEC having a
pancake structure \cite{AndEnsMat95} and alters only the dimensionality of
the resulting equations. In particular it leads to a
two-dimensional rather than a one-dimensional problem.}
Specifically, we assume that the transverse structure of the BEC
is not significantly altered by many-body interactions and is
determined as the ground state solution of the transverse potential
\begin{equation}
\hbar\omega_{\perp}v_g({\bf r}_{\perp})
= \left [-\frac{\hbar^2}{2m}
\mbox{\boldmath {$\nabla$}}_{\perp}^2
+\frac{m\omega_{\perp }^{2}} {2} {\bf r}_{\perp}^{2}\right ] 
v_g({\bf r}_\perp )  ,
\end{equation}
and we express the Hartree wave function as
\begin{equation}
\phi_{n_1,n_2}({\bf r},t) = v_g({\bf r}_{\perp})e^{-i\omega_{\perp }t}
\phi_{n_1,n_2}^\prime(z,t)  .
\end{equation}
Substituting this expression into Eq. (\ref{sys0}), projecting out
the transverse mode, and dropping the prime for simplicity in
notation, yields the coupled Gross-Pitaevskii
equations for the quasi-one-dimensional system. Introducing
the dimensionless length $\xi=z/a_z$, $a_z\equiv\sqrt{\hbar/2m\omega_z}$ being
the characteristic length associated with the longitudinal trapping potential,
and the dimensionless time $\tau=\omega_z t$, these coupled equations can
be written in the scaled form
\begin{eqnarray}
i\dot \phi_{n_1,n_2} (\xi,\tau)&=&
H_L\phi_{n_1,n_2}(\xi,\tau) +\eta |\phi_{n_1,n_2}(\xi,\tau)|^2 
\phi_{n_1,n_2}(\xi,\tau)
\nonumber\\
&+& g\left( \sqrt {n_1(n_2+1)} \phi_{n_1-1,n_2+1}+ \sqrt{(n_1+1)n_2}
\phi_{n_1+1,n_2-1}\right)  ,
\label{sys1}
\end{eqnarray}
where the Hamiltonian $H_L$ is given by
\begin{equation}
H_L = \left[-\frac{\partial^2}{\partial \xi^2}+
\frac{1}{4}\xi^{2}\right]  ,
\end{equation}
$g=\Omega_0^2/\delta\omega_z$ is the atom-field interaction energy per
photon in units of $\hbar\omega_z$, which acts as the coupling coefficient
between different states $(n_1,n_2)$, and
\begin{equation}
\eta = \frac{NV_0/{\cal V}}{\omega_z}
=\frac{NV_0}{a_z\omega_z}
\frac{\int d{\bf r}_{\perp}|v_g({\bf r}_{\perp})|^4}
{\int d{\bf r}_{\perp} |v_g({\bf r}_{\perp})|^2}  ,
\end{equation}
is the many-body interaction energy for $N$ atoms in a volume ${\cal V}$
in units of $\hbar\omega_z$.

Equations (\ref{sys1}) are the basis of the remainder of this paper.  The
transformation to dimensionless variables reveals that the key parameters
for the system are the linear coupling coefficient $g$, and the
nonlinear parameter $\eta$ describing self-phase modulation.

\section{Dressed Bose-Einstein Condenstates}
\subsection{Dressed BECs}
The dressed BECs are the eigenstates of Eqs. (\ref{sys1})
(or more generally Eqs. (\ref{sys0})), and are quantum superpositions
of states with different photon numbers $(n_1,n_2)$.  Setting
\begin{equation}
\phi_{n_1,n_2}(\xi,\tau)=e^{-i\mu \tau}\theta_{n_1,n_2}(\xi)  ,
\end{equation}
for the dressed states, we obtain
\begin{eqnarray}
\mu\theta_{n_1,n_2} (\xi)&=&
H_L\theta_{n_1,n_2} +\eta|\theta_{n_1,n_2}(\xi)|^2 \theta_{n_1,n_2}(\xi)
\nonumber\\
&+& g \left( \sqrt {n_1(n_2+1)} \theta_{n_1-1,n_2+1}+
\sqrt{(n_1+1)n_2} \theta_{n_1+1,n_2-1}\right)  ,
\label{sys2}
\end{eqnarray}
with $\mu$ the chemical potential scaled to $\hbar\omega_z$.
Admissable solutions should also be normalized according to
\begin{equation}
\int_{-\infty}^{\infty} d\xi \sum_{n_1+n_2=n}|\theta_{n_1,n_2}(\xi)|^2 = 1  .
\end{equation}
Numerical calculations are generally required to solve these equations 
for the dressed states, but simple limiting cases can be treated analytically.
\subsection{Single cavity photon, $n=1$}
The essential physics of dressed BECs can be exposed 
using the simplest case of one cavity photon, $n=1$, so only the states 
with $(1,0)$ and $(0,1)$ are relevant.  Then the coupled time-dependent 
Eqs. (\ref{sys1}) reduce to
\begin{eqnarray}
i\dot\phi_{01}(\xi,\tau) &=&
H_L\phi_{01}(\xi,\tau) + g \phi_{10}(\xi,\tau)+\eta 
|\phi_{01}(\xi,\tau)|^2\phi_{01}(\xi,\tau)  ,
\nonumber \\
i\dot\phi_{10}(\xi,\tau)&=&
H_L\phi_{10}(\xi,\tau)
+ g \phi_{01}(\xi,\tau)+\eta
|\phi_{10}(\xi,\tau)|^2\phi_{10}(\xi,\tau)  ,
\label{sys3}
\end{eqnarray}
which yields the following pair of coupled equations for the dressed states
\begin{eqnarray}
\mu\theta_{01}(\xi) &=&
H_L\theta_{01}(\xi)
+ g\theta_{10}(\xi)+\eta
|\theta_{01}(\xi)|^2\theta_{01}(\xi)  ,
\nonumber \\
\mu\theta_{10}(\xi)&=&
H_L\theta_{10}(\xi)
+ g \theta_{01}(\xi)+\eta
|\theta_{10}(\xi)|^2\theta_{10}(\xi)  .
\label{sys4}
\end{eqnarray}
These systems of equations are similar to those that appear in the theory
of multi-component condensates
\cite{GolMey97}.  However, instead of a nonlinear coupling due to cross-phase
modulation we have here linear coupling due to the exchange of photons
between cavity modes via virtual atomic transitions. 
In this respect our equations more closely resemble those describing the
{\em linear} evanescent coupling of adjacent nonlinear optical fibers
\cite{TriWabWri88}.

Equations (\ref{sys3}) are known in nonlinear physics
as the discrete self-trapping equations for a dimer
(see e.g. \cite{EilLomSco85} and references therein).
The stationary solutions for such a system were classified and their
stability was studied in Ref. \cite{EilLomSco85}
in the case $H_L=0$.  Three types of solution for the dimer were
uncovered: an in-phase solution denoted $(\uparrow\uparrow)$ for
which, in our notation, $\theta_{10}=\theta_{01}$, an out-of-phase
solution $(\uparrow\downarrow)$ with  $\theta_{10}=-\theta_{01}$,
and asymmetric solutions $(\uparrow\cdot)$ and $(\cdot\uparrow)$ with
$|\theta_{10}|\ne|\theta_{01}|$.

The same classification scheme can be employed here for the dressed
states including the Hamiltonian $H_L$ in Eqs. (\ref{sys4}).  In this
case, however, exact analytic solutions are not available for the
dressed states, but approximate solutions can be obtained for the
in-phase and out-of-phase solutions within the Thomas-Fermi
approximation in which the nonlinear interaction term dominates
over the kinetic energy term \cite{BayPet96}. 
In the Thomas-Fermi approximation we then obtain the following
dressed state solution from Eqs. (\ref{sys4})
\begin{equation}
\theta_{10}^\pm(\xi) = \frac{1}{\eta^{1/2}}
\sqrt{\mu\mp g- \frac{1}{4}\xi^2 }  ,
\label{thetaDr}
\end{equation}
when the argument of the square root is greater than or equal to zero,
and is zero otherwise. The top sign in Eq. (\ref{thetaDr}) corresponds
to the in-phase solution, and the
lower one to the out-of-phase solution.  The normalization of the wave function
then leads to the following expression for the chemical potential of the
two solutions
\begin{equation}
\mu^\pm = \pm g + \frac{1}{4}\xi_m^2  ,
\end{equation}
where $\xi_m=[3\eta/2]^{1/3}$ is the longitudinal coordinate at
which the Thomas-Fermi solution vanishes.  
Using this expression for the chemical
potential in Eq. (\ref{thetaDr}) for the dressed state solution we readily
find that the profile 
\begin{equation}
|\theta_{10}^\pm|^2 = \frac{\xi_m^2-\xi^2}{4\eta}
\label{prof}
\end{equation}
is in fact independent of $g$ and whether it is the in-phase
or out-of-phase solution.
\subsection{Elementary excitations}
The elementary excitations of the system can be found by linearizing
Eqs. (\ref{sys3}) around the dressed state solutions
\begin{eqnarray}
\phi_{10}(\xi,\tau) &=& e^{-i\mu\tau}[\theta_{10}(\xi)
+u_{10}(\xi)e^{-i\omega \tau}+v_{10}^\star(\xi) e^{i\omega \tau}],
\nonumber \\
\phi_{01}(\xi,\tau) &=& e^{-i\mu\tau}[\theta_{01}(\xi)
+u_{01}(\xi)e^{-i\omega \tau}+v_{01}^\star(\xi) e^{i\omega \tau}],
\label{uv}
\end{eqnarray}
where $u(\xi)$ and $v(\xi)$ represent small perturbations around the dressed
state with energies $\mu\pm\omega$.  In zeroth order, substitution of these 
expressions into Eq. (\ref{sys3}) results in the system of two equations
(\ref{sys4}) for the dressed states. In first-order, it leads to the system of
four equations for the linearized perturbations $u(\xi)$ and $v(\xi)$
\begin{eqnarray}
\omega u_{10}(\xi)&=&\left [ H_L
+2\eta |\theta_{10}(\xi)|^2-\mu\right]
u_{10}(\xi)+g u_{01}(\xi)+\eta \theta_{10}(\xi)^2 v_{10}(\xi),
\nonumber\\
\omega u_{01}(\xi)&=&\left [ H_L  
+2\eta |\theta_{10}(\xi)|^2-\mu\right]
u_{01}(\xi)+g u_{10}(\xi)+\eta \theta_{01}(\xi)^2 v_{01}(\xi),
\nonumber\\
\omega v_{10}(\xi)&=&\left [H_L
+2\eta |\theta_{10}(\xi)|^2-\mu\right]
v_{10}(\xi)+g v_{01}(\xi)+\eta \theta_{10}(\xi)^2u_{10}(\xi),
\nonumber\\
\omega v_{01}(\xi)&=&\left [ H_L
+2\eta |\theta_{10}(\xi)|^2-\mu\right]
v_{01}(\xi)+g v_{10}(\xi)+\eta \theta_{01}(\xi)^2 u_{01}(\xi).
\label{sysbog}
\end{eqnarray}
The normal modes of this system of coupled equations are identical
to the elementary excitations determined via the Bogoliubov method in
which the Hamiltonian for the linearized perturbations is brought into
diagonal form using a Bogoliubov transformation \cite{Bog47}.

It was shown in Ref. \cite{EilLomSco85} that for the case corresponding
to $H_L=0$ the out-of-phase solution $(\uparrow\downarrow)$
of the self-trapped equations is always
stable, while the in-phase solution $(\uparrow\uparrow)$ is stable until it
bifurcates at a condition corresponding to
$\mu^+=2g$, yielding a stable asymmetric branch and an unstable
in-phase branch.  Here we study the influence of the Hamiltonian $H_L$
on the stability of the in-phase and out-of-phase dressed-states of the system.

An exact solution for the normal modes of Eqs. (\ref{sysbog}) is not available,
to the best of our knowledge.  To proceed we therefore use the consequence
of the Thomas-Fermi approximation $\xi_m \gg 1$, which means
that the profile of the dressed state solution is broad compared to the
characteristic length scale of the trapping potential.
This  allows us to  assume
$|\theta_{n_1,n_2}^\pm(\xi)|^2\simeq|\theta_{n_1,n_2}^\pm(\xi=0)|^2
\approx \xi_m^2/4\eta$ for normal modes localized close to the center of the
trapping potential. 
With this replacement Eqs. (\ref{sysbog}) can
be conveniently solved by expanding the perturbations in terms
of eigenfunctions of the linear trapping potential
$H_L q_\nu(\xi) = (\nu+1/2)q_\nu(\xi)$
\begin{eqnarray}
u_{10}(\xi)&=&\sum_\nu b_{10}^\nu q_\nu(\xi), \mbox{\hspace{1cm}}
u_{01}(\xi)=\sum_\nu b_{01}^\nu q_\nu(\xi); \nonumber \\
v_{10}(\xi)&=&\sum_\nu c_{10}^\nu q_\nu(\xi), \mbox{\hspace{1cm}}
v_{01}(\xi)=\sum_\nu c_{01}^\nu q_\nu(\xi).
\label{anzats}
\end{eqnarray}
Substitution of these expressions into Eqs. (\ref{sysbog}) gives 
a system of linear equations for the coefficients $b^\nu$, $c^\nu$
which is straightforward to solve. For the out-of-phase dressed-state
the spectrum of the elementary excitations is
\begin {equation}
\omega=\left\{\pm\sqrt{\left(\nu+\frac{1}{2}+\frac{1}{2}\xi_m^2\right)
\left(\nu+\frac{1}{2}\right)},
\pm\sqrt{\left(\nu+\frac{1}{2}+\frac{1}{2}\xi_m^2\right)
\left(\nu+\frac{1}{2}+2g\right)}\right\}.
\label{stable}
\end{equation}
Both branches of normal modes given by Eq. (\ref{stable}) are
stable, as they are characterized by real values of $\omega$.
In contrast, the spectrum of the
normal modes for the in-phase dressed-state is
\begin{equation}
\omega=\left\{\pm\sqrt{\left(\nu+\frac{1}{2}+\frac{1}{2}\xi_m^2
-2g\right) \left(\nu+\frac{1}{2}-2g\right)},
\pm\sqrt{\left(\nu+\frac{1}{2}+\frac{1}{2}\xi_m^2\right)
\left(\nu+\frac{1}{2}\right)}\right\}.
\label{unstable}
\end{equation}
There are again two branches in the excitation spectrum, one of which,
$$
\omega=\pm\sqrt{\left(\nu+\frac{1}{2}+\frac{1}{2}\xi_m^2
-2g\right) \left(\nu+\frac{1}{2}-2g\right)}
$$
can become unstable, that is, $\omega$ can assume imaginary values.
In particular, the region of instability is defined in terms of
the index $\nu$ of the linear oscillator mode as
\begin{eqnarray}
2g-\frac{1}{2}\xi_m^2<\nu+\frac{1}{2}&<&2g 
\mbox{\hspace{1cm} for \hspace{1cm}}
g>\frac{1}{4}\xi_m^2;
\nonumber\\
\nu+\frac{1}{2}&<&2g \mbox{\hspace{1cm} for \hspace{1cm}}
g<\frac{1}{4}\xi_m^2.
\label{ineq}
\end{eqnarray}
A detailed analysis of the eigenmodes corresponding to the
elementary excitations reveals that
for the in-phase dressed state the unstable excitations
have normal modes that are $\pi$ out of phase, that is,
$u_{10}(\xi)=-u_{01}(\xi)$ and $v_{10}(\xi)=-v_{01}(\xi)$, whereas
the system is stable against symmetric perturbations,
$u_{10}(\xi)=u_{01}(\xi)$ and $v_{10}(\xi)=v_{01}(\xi)$.
This means that as the instability develops, the density profiles
of the $(1,0)$ and $(0,1)$ components should display modulations
which are $\pi$ out of phase.

From this analysis we can also estimate the index $\nu_{max}$ for
the mode of largest growth rate, that is, the most negative value of  
$$f(\nu)\equiv\omega^2=\left(\nu+\frac{1}{2}+\frac{1}{2}\xi_m^2
-2g\right) \left(\nu+\frac{1}{2}-2g\right).
$$
The requirement $f'(\nu)=0$ then leads to the condition
\begin{equation}
\nu_{max}+\frac{1}{2}=2g-\frac{1}{4}\xi_m^2  ,
\label{mumax}
\end{equation}
where the nearest integer value should be taken for $\nu_{max}$. The 
corresponding growth rate is readily found to be
$Im(\omega)=\eta |\theta_{10}(0)|^2\approx \xi_m^2/4$.
\subsection{Numerical simulation of instability}
In this section we present sample numerical simulations of the
development of the predicted instability for the in-phase dressed
state BEC.  The aim of these simulations is to validate the approximate
stability analysis of the previous section, and to
put its predictions in context using a concrete example.

We have solved the coupled Gross-Pitaevskii equations (\ref{sys3})
using a standard beam propagation technique for the initial
conditions
\begin{equation}
\phi_{10}(\xi,0)=\sqrt{1+\epsilon}\theta_{10}(\xi)  ,
\qquad
\phi_{01}(\xi,0)=\sqrt{1-\epsilon}\theta_{10}(\xi)  ,
\label{InCon}
\end{equation}
where $\theta_{10}(\xi)=\theta_{01}(\xi)$ is the in-phase ground state.
In the initial condition (\ref{InCon})
the parameter $\epsilon \ll 1$ is included to provide a slight
deviation from the exact in-phase solution.
This deviation from the exact ground state of the trapping potential
can be viewed as a wave packet of the normal modes, which triggers
any instability present in the system.

We numerically generated the in-phase ground state  solution by 
evolving the Thomas-Fermi ground state (22), which represents a
symmetric perturbation of the system and is hence stable,
until the density profile reached a steady-state.
This evolution of the initial Thomas-Fermi solution towards the
actual ground state occurs since in our numerical scheme an absorber
is placed at the spatial grid boundaries to avoid unphysical reflections 
of high spatial frequencies: The absorber removes the high spatial frequencies
present in the Thomas-Fermi solution leaving behind the actual ground
state solution for which the absorber has a negligible effect.

For the numerical simulations presented here we have set
$\eta = 80$, and thus $\xi_m = 4.93 \gg 1$, so the approximations
employed here should be valid.  Figure 1 shows the density profile
$|\phi_{10}(\xi,\tau)|^2$ for $\tau = 0$ (solid line) and
$\tau = 2$ (dashed line), and $g = 0$ so there is no coupling.
In this case the density profile remains close to the initial
profile with no sign of any density
modulations appearing, meaning that the system is stable against
density oscillations.  Figure 2 shows the evolution of the
central densities $|\phi_{10}(0,\tau)|^2$ and $|\phi_{01}(0,\tau)|^2$
for $g=7.29$.  Initially, the densities show modulations
resulting from the beating of the normal modes
excited in the initial state,
but this is followed by a region of exponential growth for one component
and decay for the other.  This is
where the excitation of largest growth rate is expected to be
dominant, and the predictions of the linear stability
analysis can be tested.  In particular, for the parameters
used here we find $\nu_{max}=8$ from Eq. (\ref{mumax}).
Figure 3 shows the density profiles $|\phi_{10}(\xi,\tau=2)|^2$ (dotted line)
and $|\phi_{01}(\xi,\tau=2)|^2$ (dashed line), along with the initial 
profile (solid line) for comparison.  Here we have plotted the densities
for $\tau=2$ in the region of exponential growth and density modulations
signaling an instability are clearly seen.  In particular, the density
oscillations of the two components are $\pi$ out of phase as predicted,
and the oscillations correspond precisely to those expected for the
most unstable mode with $\nu_{max}=8$.  

These results clearly show that the linear coupling and associated quantum
superposition of the system wave function have a large effect:  In contrast
to the stable BEC shown in Fig. 1 for $g=0$, the introduction of 
linear coupling via a single cavity photon is sufficient
to render the $N$-atom condensate spatially unstable.

The density profiles shown in Fig. 3 are appropriate if we determine which
cavity mode the photon occupies.  For example, $|\phi_{10}(\xi,\tau)|^2$
is the density given that there is one photon in mode 1 and none in mode 2.
In contrast, if no determination is made of which mode the photon occupies
then the density profile is 
$Tr_f(\rho_{atom})\equiv |\phi_{01}(\xi,\tau)|^2+|\phi_{10}(\xi,\tau)|^2$,
and this is shown in Fig. 4 where the density oscillations
remain but with sufficiently reduced contrast. Here we see the possibility for
a delayed choice experiment with a many-body system: imagine that
the system is left to evolve and then released from the trap after
which it falls under gravity and its density profile is measured.
If we measure which mode the photon occupies as the BEC is
dropping we expect the large contrast density oscillations,
whereas if we don't measure which mode the photon occupies
we expect the low contrast density oscillations.  In addition,
the decision whether to measure the cavity photon or not can
in principle be made after the BEC has left the cavity when
the BEC and field no longer interact, thus providing a
delayed choice experiment.  Care should be taken, however, that
the time at which the measurement is performed is within the
linear growth range illustrated in Fig. 2.\\
\section{Conclusion and outlook}
In this paper, we have introduced the concept of dressed condensates, 
which permit to create a coupled, multicomponent macroscopic quantum system
whose dynamics can be vastly different from that of a bare condensate. 
A number of immediate extensions of the ideas presented here can readily 
be envisioned. For example, one can easily imagine 
ways to create three-component systems, entangled condensates, etc. 
Such systems will allow one to extend many of the ideas related to measurement
theory that have been developed in recent years in quantum optics to truly 
macroscopic quantum systems. 
\acknowledgements
This work is
supported by the U.S. Office of Naval Research Contract No. 14-91-J1205,
by the National Science Foundation Grant PHY95-07639, by the
Joint Services Optics Program and by the US Army Research Office. 
The information contained in this paper
does not necessarily reflect the position or policy of the U.S. Government,
and no official endorsement should be inferred. P. M. greatfully acknowledges
an Alexander-von-Humboldt Stiftung Award which enabled his stay at
the Max-Planck Institut f\"ur Quantenoptik, during which part of this work
was carried out. He also thanks Prof. H. Walther for his warm hospitality.
\newpage
\appendix
\section{Far off-resonance single-atom \\ effective Hamiltonian}
In Section I we introduced the Hamiltonian (\ref{hamsin}) which describes
dynamics of a single-atom. This Hamiltonian conserves number of excitations
in the system, and thus its state can be represented as a linear combination
of the states with one excitation
\begin{equation}
|\psi(t)\rangle=\alpha_{00} (t)|e00\rangle+\alpha_{10}(t)|g10\rangle+
\alpha_{01}(t)|g01\rangle.
\label{st1}
\end{equation}
Equations for the coefficients $\alpha_i(t)$ follow from the Schr\"odinger
equation for the state $|\psi\rangle$ and read
(here we omit kinetic energy and trapping potential terms)
\begin{eqnarray}
i\dot\alpha_{00}(t)&=&\delta\alpha_{00}(t)+{\Omega_0}(\alpha_{10}(t)+
\alpha_{01}(t))
\nonumber\\
i\dot\alpha_{10}(t)&=&{\Omega_0}\alpha_{00}(t)
\nonumber\\
i\dot\alpha_{01}(t)&=&{\Omega_0}\alpha_{00}(t)
\label{syssin}
\end{eqnarray}
Assuming $\delta\gg 1$ the excited atomic state can be adiabatically
eliminated
$$
\alpha_{00}(t)=-\frac{{\Omega_0}}{\delta}(\alpha_{10}(t)+
\alpha_{01}(t))
$$
and the resulting equations for
the coefficients $\alpha_{01}(t)$ and $\alpha_{10}(t)$ read
\begin{eqnarray}
i\dot\alpha_{10}(t)&=&-\frac{\Omega_0^2}{\delta}(\alpha_{10}(t)+
\alpha_{01}(t))
\nonumber\\
i\dot\alpha_{01}(t)&=&-\frac{\Omega_0^2}{\delta}(\alpha_{10}(t)+
\alpha_{01}(t)).
\label{syssineff}
\end{eqnarray}
At this point it is straightforward to see that these equations follow
from the Schr\"odinger equation for the state $$|\psi_{eff}(t)\rangle=
\alpha_{10}(t)|10\rangle+ \alpha_{01}(t)|01\rangle$$ if the effective
Hamiltonian  for the system is (\ref{hamsineff}).

%\bibliography{quasi_new}
%

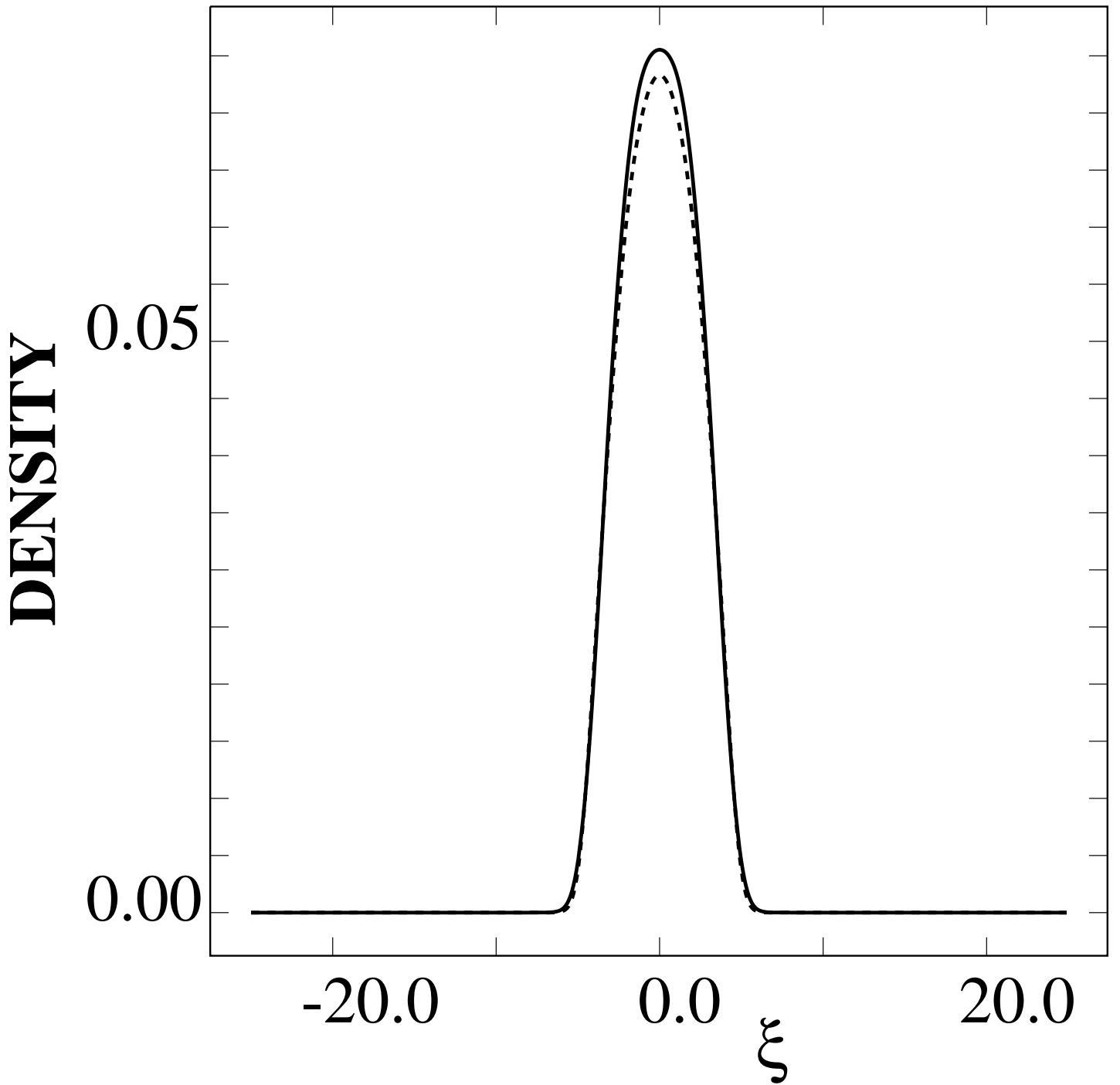
\begin{figure}
\caption{
Evolution of the density profile $|\phi_{01}(\xi,\tau)|^2$
at $\tau=2$ (dashed line) 
from the initial ground state
solution (solid line) without linear coupling $g=0$.  In this figure
and all others we set $\eta = 80$.
}
\end{figure}
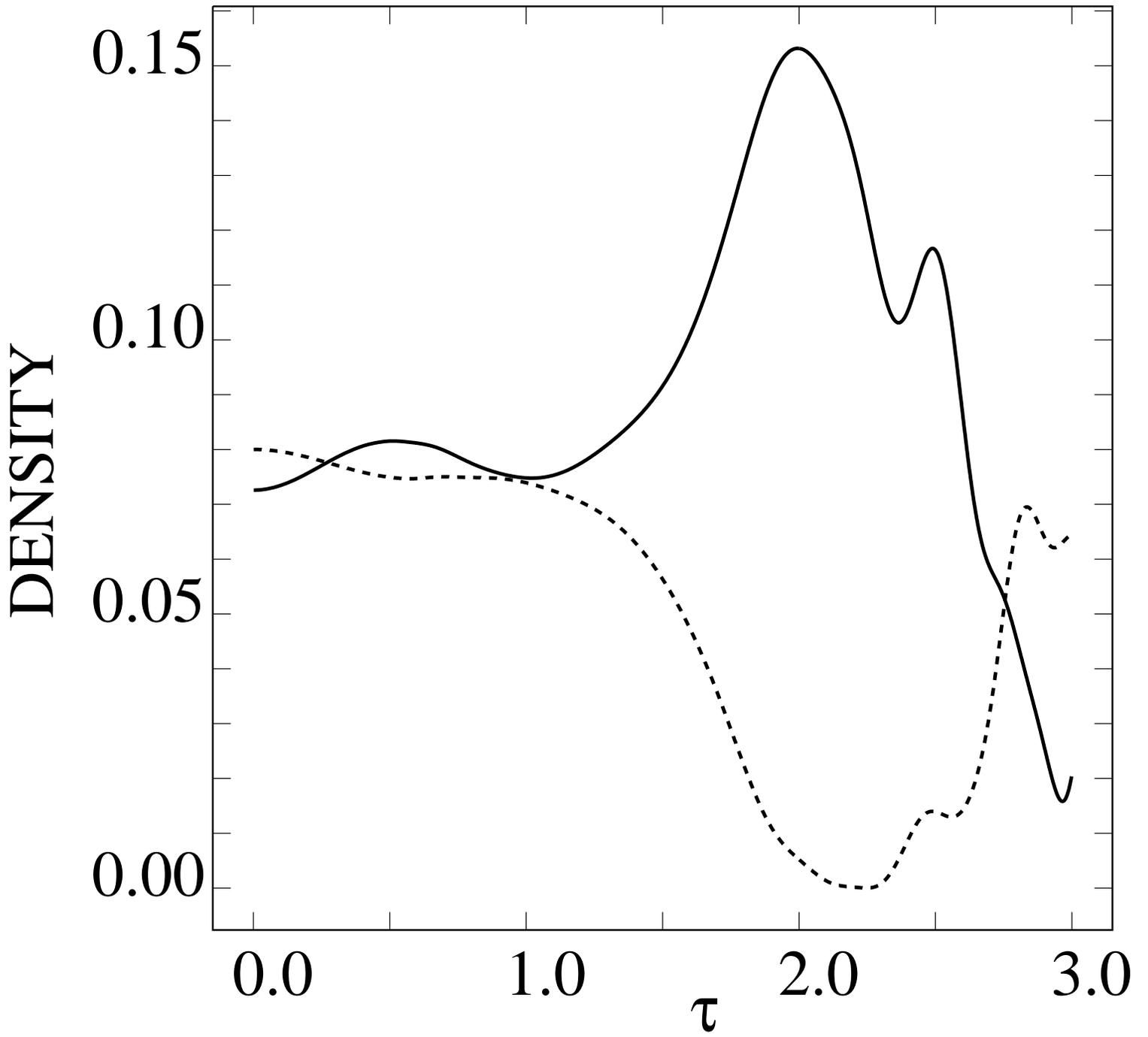
\begin{figure}
\caption{
Time dependence of the densities $|\phi_{01}(\xi,\tau)|^2$ (dashed line)
and $ |\phi_{10}(\xi,\tau)|^2$ (solid line) at the center of the trapping potential
showing the region of exponential growth for the case with linear
coupling $g=7.29$.
}
\end{figure}
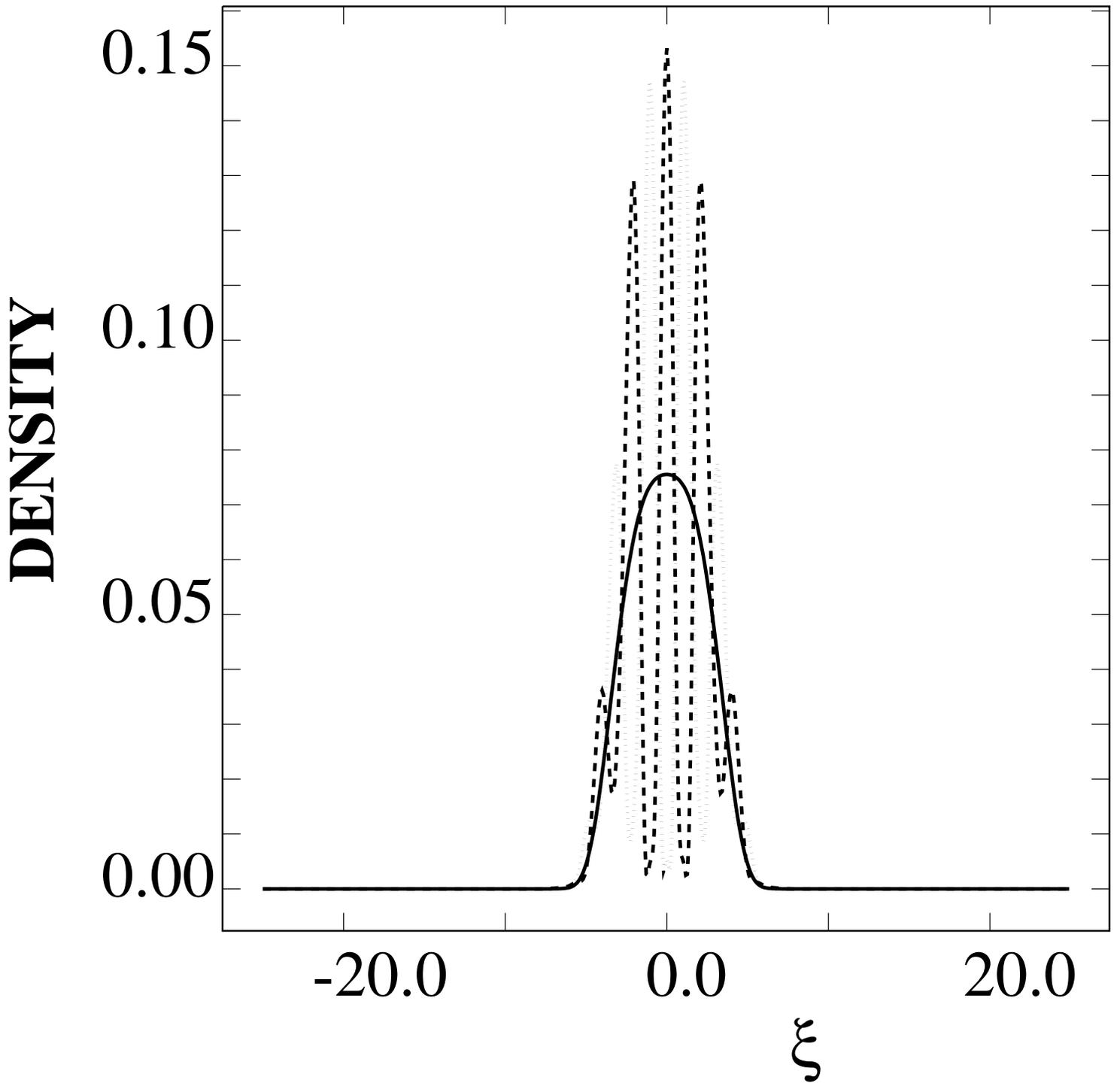
\begin{figure}
\caption{
Oscillatory pattern developed on top of the ground state profile 
in the exponential growth region $\tau=2$:
Ground state solution (solid line), $|\phi_{10}(\xi,\tau)|^2$ (dotted line), and
$|\phi_{01}(\xi,\tau)|^2$ (dashed line).
}
\end{figure}
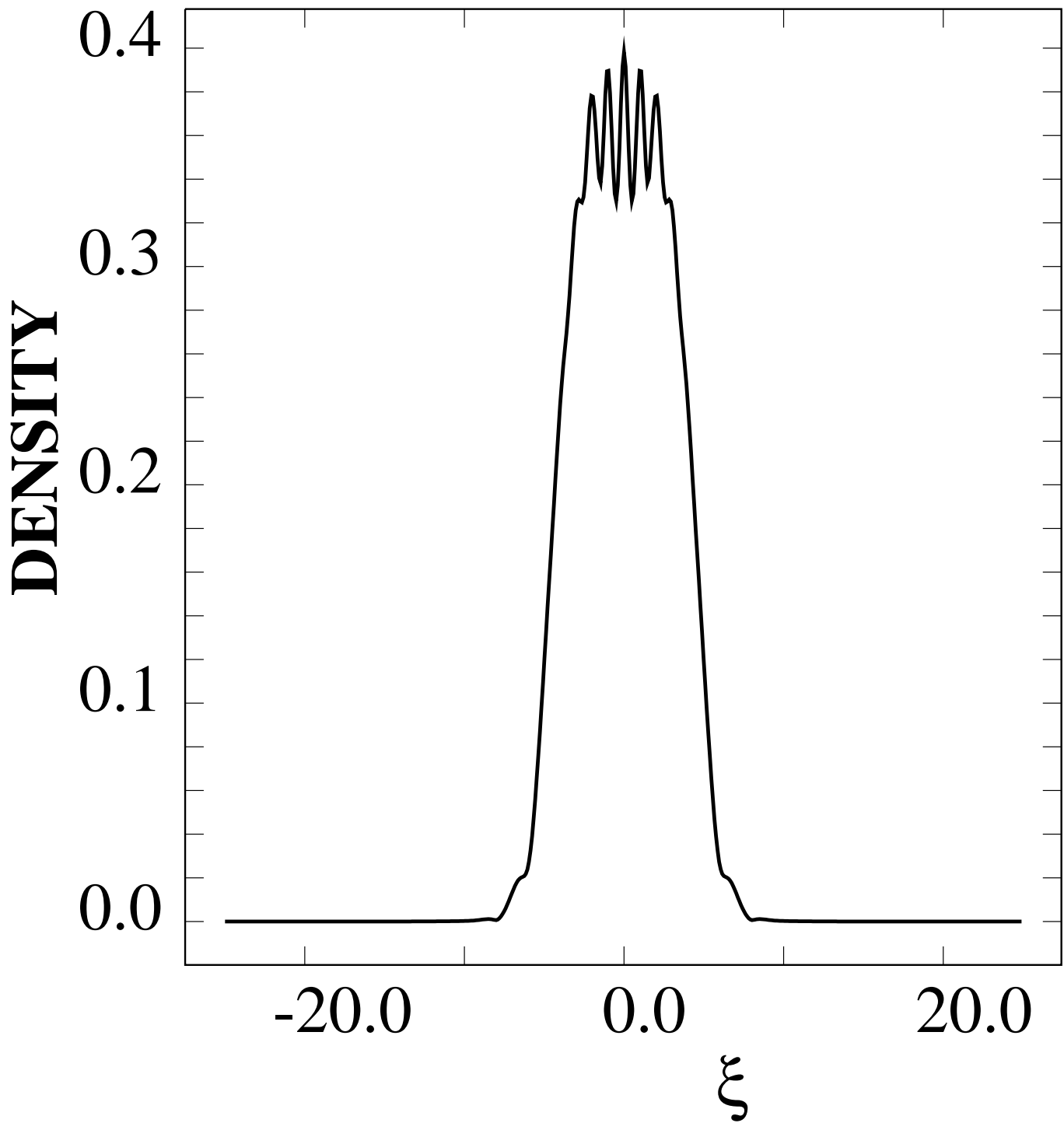
\begin{figure}
\caption{
Total density $|\phi_{10}(\xi,\tau)|^2+|\phi_{01}(\xi,\tau)|^2$ at
$\tau=2$ in the region of exponential growth for the case that no
determination is made of which mode the photon occupies.
}
\end{figure}

\end{document}